\documentstyle[prl,aps,twocolumn]{revtex}

\newcommand{\di}{\mbox{d}}

\begin{document}
\draft 
%\showrefnames
\date{\today}

\title{\Large Skyrmion in a real magnetic film.}

\author{Ar.Abanov$^{1}$ and V.L.Pokrovsky$^{1,2}$.}
\address{Department of physics, Texas A\&M University,\\ 
College Station, Texas 77843-4242, USA$^{1}$ }
\address{Landau Institute of Theoretical Physics, Moscow, Russia$^{2}$.
}

\maketitle
\begin{abstract}
Skyrmions are the magnetic defects in an ultrathin magnetic film, 
similar to the bubble domains in the thicker films.
Even weak uniaxial anisotropy determines its radius
unambiguously.
We consider the dynamics of the skyrmion decay. We show that the 
discreteness of the lattice in an isotropic 2D magnet leads  
to a slow rotation of the local magnetization in the skyrmion and, 
provided a small dissipation, to decay of the skyrmion. The radius of 
such a skyrmion as a function of time is calculated. 
We prove that uniaxial anisotropy stabilizes the skyrmion
and study the relaxation process.
\end{abstract}
\pacs{PACS numbers 75.50.Ss, 75.70.Ak, 75.70.Kw, 75.40.Gb}

%\pagebreak

%\section{Skyrmion and Anisotropy.}
\par
Skyrmions are topological excitations of 2D magnet or ultrathin
magnetic films similar to the well-known bubble domains in the industrial
magnetic memory materials \cite{bubble}. The former play an important
role as letters in magnetic records. A natural question arises whether the 
skyrmions can be utilized in a similar way. To answer this question one 
must adjust the existing theory to a realistic conditions of real films 
with their anisotropy, defects, dissipation, and discreteness. It is 
necessary also to consider how the skyrmion can be created and destroyed,
i.e. the skyrmion dynamics, ignored in the previous studies. In this
paper we address these problems.
\par
The Skyrmion was first discovered by Skyrm \cite{Skyrm} who considered 
it as a localized solution in a model of the nuclear matter. 
Belavin and Polyakov (BP) \cite{Polyakov} have shown that the 
skyrmion is a topologically nontrivial minimum of energy for the so-called 
$\vec{\bf n}$-field, the classical continuous limit of the Heisenberg
model. It realizes the mapping of the plane in which the spins 
are placed onto sphere of the order parameter with the
degree of mapping $1$.
\par
Though the $\vec{\bf n}$-field model was inspired by the studies of magnetic 
films, the first more or less direct observation of the skyrmions was made
on the 2D electron layers under the QHE conditions \cite{QHE} following an
earlier theoretical predictions \cite{Sondhi}. An indirect observation
of skyrmion effects in quasi-2D magnets was reported by F.Waldner \cite{W}
who found the skyrmion energy from the heat capacity measurements 
in a good agreement with the theoretical prediction by BP. This experimental
observation is highly nontrivial since the real magnets are at least
weakly anisotropic. Due to the existence of Goldstone modes in the 
Heisenberg magnet even very weak anisotropy can change the excitations
crucially.
\par
Let us first approach the problem with simple dimensionality arguments.
The skyrmion is a static excitation of the homogeneous ferromagnetic state
localized in a circle of the radius $R$. From the dimensionality 
consideration and from the BP results the skyrmion energy does not depend
on its size and is equal to $4\pi J|m|$ where $m$ is the degree of mapping.
Hence, the exchange energy 
$1/2\int J(\nabla {\bf S})^{2}\di^{2}x$ 
of a skyrmion
does not depend on its radius $R$. The anisotropy
energy $1/2\int \lambda S_{z}^{2}\di ^{2}x$ of the skyrmion is proportional 
to $R^{2}$ and decreases together with $R$.
Therefore the exchange energy of the fourth order in the derivatives 
$1/2\int \kappa (\Delta {\bf S})^{2}\di ^{2}x$ is crucial for the skyrmion. 
It stabilizes the skyrmion if $\kappa >0$.
Now the total energy of the skyrmion
depends on its radius and has the minimum at $R\sim (\kappa /\lambda)^{1/4}$.
One can estimate $\kappa $ by the order of magnitude as $Ja^{2}$ where $a$ is
the lattice constant. Then $R\sim (al)^{1/2}\ll l_{\lambda }$
where $l_{\lambda }=\sqrt{J/\lambda }$
is the domain wall width. The energy
of such a skyrmion differs by about 
$\lambda R^{2}\sim \sqrt{\kappa\lambda }$ from the classical skyrmion 
energy $4\pi J\gg \sqrt{\kappa \lambda }$. 
\par
Let us consider the situation more closely. The classical two-dimensional
Heisenberg exchange ferromagnet in continuous approximation is described 
by the Hamiltonian:
\begin{equation}\label{Hamiltonian-0}
\mathsc{H}_{0}=\frac{1}{2}\int J(\nabla {\bf S})^{2}\di^{2}x
\end{equation}
with the constraint on the vector field ${\bf S}({\bf r})$: 
${\bf S}^{2}({\bf r})=1$.
An obvious minimum of such a Hamiltonian is the homogeneous ferromagnetic 
configuration in which all the spins are parallel ${\bf S}({\bf r})=const$.
The simplest topologically non-trivial minimum of the 
Hamiltonian (\ref{Hamiltonian-0}) is given by the Skyrm solution: 
\begin{eqnarray}
S_{0x}&=&\frac{2Rr}{R^{2}+r^{2}}\cos (\phi +\psi) \nonumber \\
S_{0y}&=&\frac{2Rr}{R^{2}+r^{2}}\sin (\phi +\psi) \nonumber \\
S_{0z}&=&\frac{R^{2}-r^{2}}{R^{2}+r^{2}}. \label{skyrmion}
\end{eqnarray}
It describes a skyrmion of radius $R$ with the center placed in the 
origin. The observation point is indicated by 
the polar coordinates $r$ and $\phi$; $\psi$ is an arbitrary angle.
We have already mentioned that the energy $\mathsc{E}$ of the 
skyrmion (\ref{skyrmion}) does not depend on its radius $R$ 
It also does not depend on the angle $\psi$. 
\par
Any static distribution of magnetization ${\bf  S}({\bf r})$ satisfies the 
equilibrium equation:
\begin{equation}\label{static}
\frac{\delta \mathsc{H}}{\delta {\bf S}({\bf r})}=
{\bf  S}({\bf r})\left({\bf  S}({\bf r})\cdot 
\frac{\delta \mathsc{H}}{\delta {\bf S}({\bf r})}\right).
\end{equation}
The r.-h.-s. of eqn. (\ref{static}) is added as a Lagrangian factor which 
ensures that ${\bf S}^{2}({\bf r})=1$ at any point ${\bf r}$. For 
$\mathsc{H}=\mathsc{H}_{0}+\mathsc{H}_{1}$ where $\mathsc{H}_{1}$ is a
perturbation we look for a solution in a form 
${\bf S}={\bf S}_{0}+{\bf S}_{1}$, where ${\bf S}_{0}({\bf  r})$ is determined
by eqn (\ref{skyrmion}) and ${\bf S}_{1}({\bf r})$ is perpendicular 
to ${\bf S}_{0}$ and satisfies the linearized inhomogeneous equation:
\begin{equation}\label{linearized}
\hat{\mathsc{K}}{\bf S}_{1}({\bf r})=
\left[\frac{\delta \mathsc{H}_{1}}{\delta {\bf S}({\bf  r})} -
{\bf S}_{0}({\bf r})\left({\bf S}_{0}\cdot 
\frac{\delta \mathsc{H}_{1}}
{\delta {\bf S}} \right)\right]_{{\bf S}={\bf S}_{0}}
\end{equation}
where the tensor kernel $K _{ij}({\bf r},{\bf r}')$ of the linear
operator $\hat{\mathsc{K}}$ is given by
\begin{eqnarray}
&K _{ij}&({\bf r},{\bf r}')
=-J\delta ({\bf r}-{\bf r}')\times      \nonumber \\
&\bigl(\delta _{ij}&\Delta _{{\bf r}'}-
\delta _{ij}\left({\bf  S}_{0}\cdot \Delta {\bf S}_{0} \right)-
S_{0i}S_{0j}\Delta _{{\bf r}'} 
\bigr).     \nonumber
\end{eqnarray}
\par
The derivatives $\partial {\bf S}_{0}/\partial R$ and 
$\partial {\bf S}_{0}/\partial \psi$ are the zero modes and, hence, 
satisfy the homogeneous equation
$\hat{\mathsc{K}}\partial {\bf S}_{0}/\partial R=
\hat{\mathsc{K}}\partial {\bf S}_{0}/\partial \psi=0$. Therefore
the right hand side of eqn. (\ref{linearized}) must be orthogonal
to the vectors $\partial {\bf S}_{0}/\partial R$ and 
$\partial {\bf S}_{0}/\partial \psi$. The two orthogonality conditions
allow us to determine both $R$ and $\psi$ fixed by the small perturbation
$\mathsc{H}_{1}$. Considering a special perturbation Hamiltonian
\begin{equation}\label{perturbat}
\mathsc{H}_{1}=\frac{1}{2}\int \left(\kappa (\Delta {\bf S})^{2}+
\lambda (1-S_{z}^{2}) \right)\di^{2}x,
\end{equation}
we find from the orthogonality condition:
\begin{equation}\label{radius}
R=R_{0}=\left(\frac{8\kappa }{3\lambda L} \right)^{1/4}
\end{equation}
where $L=\log[(3\lambda /8\kappa )^{1/4}l_{\lambda }]$. 
Logarithm in eqn. (\ref{radius}) comes from
the divergent integral $\int(1-S_{0z}^{2})\partial S_{0z}/\partial R\di ^{2}x$.
It was cut off at the radius $r=l_{\lambda }$ at 
which the perturbation theory fails.
Due to the axial symmetry $\psi$ remains zero mode. Note that the skyrmion 
does not exist for $\kappa <0$. There is no general argument in favor of 
positive $\kappa $, but for standard discrete Heisenberg model $\kappa $ is
positive. Now we are in position to consider the dynamics of the skyrmion.
\par 
The dynamics of the unit vector field ${\bf S}({\bf r},t)$ is given by the 
Landau-Lifshitz equation \cite{LL}:
\begin{eqnarray}
\dot{\bf S}({\bf  r},t)&=&-g{\bf S}({\bf r},t)\times 
\frac{\delta \mathsc{H}[{\bf S}]}{\delta {\bf S}({\bf r},t)}+ \nonumber \\
\nu {\bf S}({\bf r},t)&\times&\left[{\bf S}({\bf r},t)\times 
\frac{\delta \mathsc{H}[{\bf S}]}
{\delta {\bf S}({\bf r},t)}\right]. \label{Landau-Lifshic-diss}
\end{eqnarray}
It can be checked that at $\nu =0$ the equation of motion 
(\ref{Landau-Lifshic-diss}) conserves the magnetization
and energy of the field ${\bf S}({\bf  r})$ as well as the local 
constraint ${\bf S}^{2}({\bf r})=1$. A small dissipation term
at $\nu \not= 0$ allows for the relaxation processes.
\par
We consider only the slow skyrmion dynamics.
It means that we present again the Hamiltonian $\mathsc{H}$ as a sum
$\mathsc{H}_{0}+\mathsc{H}_{1}$, where $\mathsc{H}_{1}$ is a small
perturbation to the exchange Hamiltonian $\mathsc{H}_{0}$. We are looking
for a solution in the form ${\bf S}={\bf S}_{0}+{\bf S}_{1}$, 
${\bf S}_{0}\cdot  {\bf S}_{1}=0$, where $|{\bf S}_{1}|\ll |{\bf S}_{0}|$ and
${\bf S}_{0}({\bf r};R(t),\psi(t))$ is the standard skyrmion solution
(\ref{skyrmion}) with the parameters $R$ and $\psi$ slowly varying in time.
We also consider the dissipation  and the term $\dot{\bf S}$ as a perturbation.
Substituting only ${\bf S}_{0}({\bf r};R(t),\psi(t))$ in the perturbation 
terms in eqn. (\ref{Landau-Lifshic-diss}) and requiring the orthogonality
of the perturbation terms to both zero modes $\partial {\bf S}_{0}/\partial R$
and $\partial {\bf S}_{0}/\partial \psi$, we obtain equations of motion for
$R$ and $\psi$:
\begin{eqnarray}
\frac{\dot{R}}{R}&=&-\frac{\nu }{g}\omega \label{dissipation}\\
-&\omega& R^{2}\ln \frac{\tilde{R}^{2}}{R^{2}}+
g\kappa \frac{16}{3}\frac{1}{R^{2}}-
g\lambda \ln \frac{\tilde{R}^{2}}{R^{2}}=0 \label{rotation}
\end{eqnarray}
where $\omega =\dot{\psi}$ and $\tilde{R}$ is a scale at which  
the perturbation expansion breaks down, i.e.
$|{\bf S}_{0}(r=\tilde{R})|\approx |{\bf S}_{1}(r=\tilde{R})|$.
Eqns (\ref{dissipation}) and (\ref{rotation}) have a fixed point with
$\omega =0$ and $R=R_{0}$, where $R_{0}$ is determined by eqn. (\ref{radius}).
\par 
If $R$ deviates slightly from $R_{0}$ so that $\Delta R=R-R_{0}$ is still
small enough, one can put $\tilde{R}=l_{\lambda }$ into 
(\ref{rotation}) and obtain $\omega =4\lambda g \Delta R/R$.
Using (\ref{dissipation}), we find:
\begin{equation}\label{damping}
\Delta \dot{R}=-4\lambda \nu \Delta R.
\end{equation}
``Small enough'' $\Delta R$ means that we still can use 
$\tilde{R}\approx l_{\lambda }$. However, if $\Delta R$
is large, the cut-off scale is determined by the finite frequency 
$\tilde{R}\approx l_{\omega}= \sqrt{gJ/|\omega |}$.
Eqn. (\ref{damping}) is valid if $l_{\lambda }<l_{\omega }$ 
or $\Delta R/R< 1/4$.
\par
In the opposite case 
$l_{\omega}\ll l_{\lambda }$
and $R\gg R_{0}$ one can neglect the second term in (\ref{rotation}).
Then equations of motion (\ref{dissipation}, \ref{rotation}) read as follows
\begin{equation}\label{big}
\omega = \lambda g;\,\,\,   \dot{R}=-\nu \lambda R.
\end{equation}
For $l_{\omega }\ll l_{\lambda }$ and $R\ll R_{0}$ one can drop 
the last term in (\ref{rotation}). Then:
\begin{equation}\label{small}
\omega =-\frac{16}{3}\frac{\kappa g}{R^{4}}\frac{1}
{\ln \frac{3R^{2}J}{16\kappa }};\,\,\,\,          
R^{3}\dot{R}=\frac{16}{3}\kappa \nu \frac{1}
{\ln \frac{3R^{2}J}{16\kappa }}.          
\end{equation}
\par
Thus, the easy-axis anisotropy together with the fourth-order exchange
term fix the radius (\ref{radius}) of the skyrmion if $\kappa >0$.
In the opposite case $\kappa <0$ there is no stable configuration unless 
the higher order exchange interaction is taken into account. We also
have shown that the skyrmion reaches its equilibrium radius with 
the characteristic time $t_{r}=(\nu \lambda )^{-1}$. 
\par 
Another property of a real 
film which should be taken into account is the lattice discreteness.
We have mentioned earlier that in the continuous model 
the skyrmion is a topological excitation and as such cannot dissipate.
However, in the discrete lattice the continuity of the field 
${\bf S}({\bf r})$ is lost and the very notion of the topological excitation
becomes inconsistent. Therefore, the skyrmion configuration in 
the discrete lattice is unstable. Moreover it is sufficient to
remove one plaquette in the center of the skyrmion to make it unstable.
(see for example \cite{Haldane}). We will imitate the 
discreteness effect by considering a hole in the center of the skyrmion. 
In the picture due to BP 
the skyrmion is described by a meromorphic function which has a pole
in the center of the skyrmion. This pole cannot be rid off by any 
continuous change of the field ${\bf S}$.
To allow a skyrmion to dissipate we should punch a small hole in its center.
\par 
Let us first consider the skyrmion without anisotropy.
``Punching a small hole in the center'' means the substitution
$J\rightarrow J\theta (r-r_{0})$ in eqn. (\ref{Hamiltonian-0}) where
$r_{0}\ll R$ is the radius of the hole and $\theta (x)$ is the
step function: $\theta (x)=0$ for $x<0$ and $\theta (x)=1$ for $x>0$. 
Hence, the perturbation
$\mathsc{H}_{1}$ is given by:
\begin{equation}\label{Ham1-hole}
\tilde{\mathsc{H}}_{1}=-J\int \theta (r_{0}-r) (\nabla {\bf S})^{2}
\end{equation}
Employing the orthogonality condition of the right hand side of 
the linearized eqn. (\ref{Landau-Lifshic-diss})
to the zero modes and neglecting all the terms
of the higher order in $r_{0}$, one finds that eqn. (\ref{dissipation})
still holds, but eqn. (\ref{rotation}) must be replaced by: 
\begin{equation}\label{rotation-hole}
\omega \ln \frac{\tilde{R}^{2}}{R^{2}}=g\frac{Jr_{0}^{2}}{R^{4}},
\end{equation}
where $\tilde{R}\approx l_{\omega }=\sqrt{gJ/|\omega|}$ 
is the scale where the perturbation scheme
breaks down. With the logarithmic accuracy one can write 
$\ln (\tilde{R}^{2}/R^{2})=\ln (R^{2}/r_{0}^{2})$.
Finally, substituting $\omega $ from (\ref{rotation-hole})
into (\ref{dissipation}), we find:
\begin{equation}\label{dissipation-hole}
R^{3}\dot{R}=-\nu\frac{J r^{2}_{0}}{\ln \frac{R^{2}}{r^{2}_{0}}}
\end{equation}
From this equation one can conclude that the skyrmion's life-time is
roughly proportional to its radius in the fourth power. 
\par
Returning to the field with the easy-axis anisotropy and 
the fourth-order term, let us introduce again a hole in the center. 
In this situation the perturbation is given
by the sum of the two terms (\ref{perturbat}) and (\ref{Ham1-hole}).
Using the same orthogonality trick one gets:
\begin{equation}\label{hole-log-anisotr}
-2\frac{\omega }{g}R^{2}\ln \frac{\tilde{R}^{2}}{R^{2}}-\frac{32}{3}
\frac{\kappa }{R^{2}}+2R^{2}\lambda \ln \frac{\tilde{R}^{2}}{R^{2}}=
-4J\frac{r_{0}^{2}}{R^{2}}.
\end{equation}
After substitution 
$\kappa \rightarrow \tilde{\kappa}=\kappa -(3/8) Jr_{0}$ eqn.
(\ref{hole-log-anisotr}) acquires the same form as 
eqn. (\ref{rotation}). It means that 
as long as $\kappa >(3/8) Jr_{0}^{2}$
the skyrmion is stable and its radius is defined by eqn. (\ref{radius}) with
$\tilde{\kappa}$ instead of $\kappa$. Equations (\ref{damping}),
(\ref{big}) and (\ref{small}) are valid as well after the same substitution.
In the case $\kappa < (3/8) Jr_{0}^{2}$, however, the effective 
$\tilde{\kappa}$ is negative and the stable skyrmion exists no longer.
%Then for the radius $a\gg 32|\tilde{\kappa}|/
%3\lambda \ln (3J^{2}/32\kappa \lambda )$ the equation (\ref{big}) still
%holds while for the opposite case the equation (\ref{small}) must be
%changed.
%\begin{eqnarray}
%\omega =\frac{16}{3}\frac{|\tilde{\kappa} |g}{a^{4}}\frac{1}
%{\ln \frac{3a^{2}J}{16|\tilde{\kappa}|}}            \nonumber \\
%a^{3}\dot{a}=-\frac{16}{3}|\tilde{\kappa}| \nu \frac{1}
%{\ln \frac{3a^{2}J}{16|\tilde{\kappa}|}}              \label{hole-small}
%\end{eqnarray}
%We see that if the hole is big enough then the skyrmion decays.
\par 
In conclusion, we have shown that 
in a real ultrathin ferromagnetic film with 
easy-axis anisotropy the skyrmion known for the isotropic model
still exists, but it
acquires a definite radius $R_{0}$ given by eqn. (\ref{radius}). 
By the order of magnitude $R_{0}\sim \sqrt{a l_{\lambda }}\ll l_{\lambda }$ 
where $l_{\lambda }$ is the domain wall width. 
Once we made a domain with the reversed magnetization
in a ferromagnetic film it shrinks down to the size 
$R_{0}\sim 1nm$. The
discreteness of the lattice in the isotropic
model leads to a finite skyrmion life-time which is roughly proportional
to the fourth power of its radius. However, 
anisotropy together with the higher order exchange interaction 
stabilizes the skyrmion. At finite temperature it can decay 
through an instanton configuration. Our results allow to understand 
why the activation energy found by Waldner is so close to $4\pi J$:
the difference is expected to be of the 
relative order $a/l_{\lambda }\sim 10^{-2}$.
The detailed thermodynamics of skyrmions will be published elsewhere.
\par
This work has been supported by the NSF grant DMR-9705812 and
by the DOE grant DE-FGO3-96ER45598.

\end{document}